\documentstyle[12pt,epsf,epsfig]{ioplppt}          
\begin{document}
\jl{1}

\title{Statistical Physics of Irregular  Low-Density Parity-Check Codes}
[Statistical Physics of Irregular Codes]

\author{Renato Vicente \dag \ftnote{3}{E-mail: vicenter@aston.ac.uk},  
        David Saad \dag\ and Yoshyiuki Kabashima \ddag }

\address{\dag\ The Neural Computing Research Group, Aston University, Birmingham B4 7ET, UK}

\address{\ddag\ Department of Comptational  Intelligence and  System Science, Tokyo Institute of Technology, Yokohama 2268502, Japan}

\begin{abstract}
Low-density parity-check codes with irregular constructions 
have been recently shown  to outperform the most advanced 
error-correcting codes to date. 
In this paper we apply methods of statistical physics to study  
the typical properties of simple  irregular codes.
 We use  the replica method to  find a phase  transition which coincides 
with Shannon's coding bound when appropriate parameters are chosen.
 The decoding by belief propagation 
is also studied using statistical physics arguments; the theoretical 
solutions 
obtained are in  good agreement with simulations.
We compare the performance of irregular  with that of regular 
codes and discuss the factors that contribute to the improvement in
 performance.
\end{abstract}



\section{Introduction}
Error-correction mechanisms are essential for preventing loss of information
in transmissions through noisy environments. They are of increasing technological importance with 
applications ranging from high capacity storage media to  satellite 
communication. The surprising fact that 
error-free communication is possible if the information is encoded  to include a minimum amount of redundancy was discovered by Shannon in 1948 
\cite{shannon}. 
Shannon proved that a message encoded at rates $R$ (message information content/code-word length) up to the channel capacity $C$ can be  decoded
with   vanishing average error probability $P_E\rightarrow 0$ as the length 
of the message increases $M\rightarrow \infty$. This theorem was then progressively refined 
by Gallager and others (see \cite{viterbi} and references therein) to say that
the average over messages and codes of the error probability is bounded
by
\begin{equation}
P_E < \mbox{e}^{-M E(R)},
\end{equation} 
where $E(R)$ is the error exponent that is greater than zero for rates up
to the channel capacity $C$.

These proofs were presented in a non-constructive form by assuming
encoding processes by ensembles of unstructured random codes and 
 impractical decoding methods like  
maximum likelihood  or  typical set decoding \cite{cover}. 
No encoding-decoding scheme that is practical and attains 
the coding bound  has been found to date.

The most successful code in use to date is the 
 Turbo code \cite{turbo}. However, the current performance record 
is owned by an irregular low-density parity check code (LDPC), more specifically
an irregular Gallager code \cite{davey2}
\footnote {See {\em http://www331.jpl.nasa.gov/public/JPLtcodes.html}
 for JPL's  `` imperfectness'' contest.}.
This code was first proposed by Gallager in 1962 \cite{gallager0,gallager1}, 
and were all but  forgotten soon after due to technical limitations of the 
time. Recently a variation of the original proposal by Gallager named MN code 
has been proposed by MacKay and Neal \cite{mackay0,mackay1}; they showed that 
this code has good performance, what attracted renewed interest to LDPCs. 
Since then LDPCs  have been   reconsidered in a variety of architectures
 \cite{mackay2,luby}. Some of which reported close to optimal performance 
\cite{urbanke,IKS}.

Representing  a message by a binary vector $\mbox{\boldmath $\xi$}\in\{0,1\}^N$, the LDPC encoding process  consists of  producing the binary vector 
$\mbox{\boldmath $t$}\in\{0,1\}^M$  defined by  
$\mbox{\boldmath $t$}=\mbox{\boldmath $G^T$}\mbox{\boldmath $s$}\;\mbox{(mod 2)}$, where
all  operations are performed in the field  $\{0,1\}$ and are indicated by $\mbox{(mod 2)}$ and 
 $\mbox{\boldmath $G^T$}$ is a  $M \times N$ generator matrix.  The transmission  is then corrupted by noise, that we assume to be a binary vector $\mbox{\boldmath $\zeta$}\in\{0,1\}^M$,    
and the received vector takes the form $\mbox{\boldmath $r$}=\mbox{\boldmath $G^T$}\mbox{\boldmath $\xi$}+\mbox{\boldmath $\zeta$}\mbox{ (mod 2)} $. The decoding process is performed by  applying  a suitable  parity-check  matrix to the received message to produce
the {\it syndrome}  vector  $\mbox{\boldmath $z$}=\mbox{\boldmath $A$}\mbox{\boldmath $r$}\;\mbox{(mod 2)}$.
The parity-check  matrix $\mbox{\boldmath $A$}$  defines the code structure and  can be represented 
by a bipartite undirected graph with check   and bit  nodes. This gives rise to the classification of LDPCs to regular (those forming  regular graphs)
 and irregular codes.

The parity-check matrix for  Gallager  codes  is a concatenation
 $\mbox{\boldmath $A$}= [\mbox{\boldmath $C_1$}\mid\mbox{\boldmath $C_2$}]$
of two very sparse matrices, with 
 $\mbox{\boldmath $C_2$}$ (of dimensionality $(M-N)\times (M-N)$) being invertible and
the rectangular matrix  $\mbox{\boldmath $C_1$}$ of dimensionality  $(M-N)\times N$. 
The generator matrix of a Gallager code is $\mbox{\boldmath $G$}=
[\mbox{\boldmath $I$}\mid\mbox{\boldmath $C_2^{-1}$}\mbox{\boldmath $C_1$}]
\mbox{ (mod 2)}$, where $\mbox{\boldmath $I$}$ is the  $N\times N$ identity matrix, implying  that $\mbox{\boldmath $A$}\mbox{\boldmath $G^T$}
\mbox{ (mod 2)} =0$ and that the message itself is set as the first $N$ bits in the transmission. The syndrome vector is then   $\mbox{\boldmath $z$}=\mbox{\boldmath $A$}\mbox{\boldmath $r$}=\mbox{\boldmath $A$}\mbox{\boldmath $\zeta$}\mbox{ (mod 2)}$ from which the noise can be estimated and subtracted from the received message.
For a MN code the generator matrix has the form  $\mbox{\boldmath $G^T$}= 
\mbox{\boldmath $C_n^{-1}$}\mbox{\boldmath $C_s$}\mbox{ (mod 2)} $, where 
$\mbox{\boldmath $C_n$}$ is an $M\times M$ invertible matrix and 
$\mbox{\boldmath $C_s$}$ is $M\times N$. The 
matrix applied by the decoder is given by $\mbox{\boldmath $C_n$}$ producing $\mbox{\boldmath $z$}=\mbox{\boldmath $C_n$}\mbox{\boldmath $r$}=\mbox{\boldmath $C_s$}\mbox{\boldmath $\xi$}+\mbox{\boldmath $C_n$}\mbox{\boldmath $\zeta$}\mbox{ (mod 2)} $,
 from which  the most probable  message vector can be predicted.

Although Gallager  and MN codes can be analysed  by  the same methods of
information theory  \cite{mackay1}, they represent different physical 
systems with different properties. In this paper we will restrict
the analysis to irregular MN codes, the analysis of Gallager codes  will appear  elsewhere.

Statistical physics has first been applied to the analysis of error-correcting
codes in the seminal work of Sourlas \cite{sourlas} which has been recently 
extended to the case of finite code rates \cite{KS,VSK}. 
Similar methods have been recently applied to the case of Turbo codes 
\cite{andrea} and regular MN codes \cite{KMS,MKSV}, providing a detailed 
description of the system's phases and capabilities for various parameter
choices. Here we analyse irregular MN codes using the  standard replica 
calculation to find a free energy that is a measure of 
the likelihood of typical solutions to the decoding problem,  given 
an ensemble of code matrices $\mbox{\boldmath $C_s$}$ and 
$\mbox{\boldmath $C_n$}$ ({\it code construction}), channel and message
models ({\it noise level} and {\it message bias}).

 We show that three types of solutions emerge depending on the parameters
provided: successful errorless decoding (number of incorrect bits less than 
${\cal O}(N)$),  imperfect decoding (number of incorrect bits of order $N$)
 and complete failure (number of correct bits less than 
${\cal O}(N)$).  
We  also show, as in \cite{KMS,MKSV}, that the line  separating errorless
 and complete failure phases can  coincide with the coding limit; this
 fact itself  is not particularly surprising  as the statistical
 physics analysis relies
on the same kind of arguments used in the original coding bounds, using
averages over ensembles of codes and maximum likelihood decoding. The main 
 difference here is that the matrices in the ensemble have some structure. 

The statistical physics approach can be regarded as complementary to 
that of information theory; it enables one to attain a more complete 
picture by analysing the decoding problem in the infinite message limit
and by  looking at global properties of the free energy. It allows for a 
transparent analysis of the possible performance of different codes 
 characterised by  different choices of construction parameters, and has 
 already resulted in new practical high performance codes \cite{IKS}.

In this  framework,  Bayes-optimal decoding generally  corresponds to finding the global minimum of a TAP free energy \cite{tap,plefka} which is very costly if the landscape has multiple  local minima. A practical decoding algorithm that has been used  in LDPCs is the  scheme known as  belief propagation, broadly used by the Bayesian inference community \cite{pearl,cheng}.  
Belief propagation is equivalent to solving iteratively a set of coupled
equations for finding extrema (local or global) of the TAP free energy
\cite{VSK,MKSV,KS0}. This method is very sensitive to the presence of 
local minima and can be  easily trapped in sub-optimal solutions.  

In this paper we study the dependence of the free energy surface on the 
noise level and the message bias; this allows us to study the solutions 
which exist in each one of the cases and to detect the emergence of 
suboptimal solutions that will interfere in the practical decoding dynamics.

This paper is organised as follows: Section 2 presents irregular MN codes,
 while the statistical physics analysis is outlined in Section 3; 
the relations between the belief propagation approach  and statistical 
physics  are discussed in Section 4 and employed to examine  the decoding 
performance in Sections 5 and 6. Concluding remarks are given in Section 7.


\section{Irregular MN codes}

Although the best irregular LDPCs found so far are defined in $q$-ary
alphabets \cite{davey}, we will restrict the current analysis to the  
binary alphabet $\{0,1\}$.

We suppose that the binary messages  $\mbox{\boldmath $S$}$ comprise
 independent bits sampled from the prior distribution 
$P(S)=(1-p)\,\delta(S)+p\,\delta(S-1)$, where $\delta(S)$ stands for the 
Dirac's delta distribution.
We also assume  a simple memoryless Binary Symmetric Channel (BSC) with
binary  vectors  $\mbox{\boldmath $\tau$}$ having independent components
 sampled from a similar prior distribution of the 
form $P(\tau)=(1-f)\,\delta(\tau)+ f\,\delta(\tau-1)$. From now
on we will reserve the symbols $\mbox{\boldmath $\xi$}$ and 
$\mbox{\boldmath $\zeta$}$ for the actual message and noise, using
$\mbox{\boldmath $S$}$ and $\mbox{\boldmath $\tau$}$ for denoting
random variables in the message and noise models.

The goal is then to find the Bayes-optimal estimate 
 $\widehat{S}_j=\mbox{argmax}_{S_j} \mbox{Tr}_{S_{i\neq j},\mbox{\boldmath $\tau$}} P(\mbox{\boldmath $S$},\mbox{\boldmath $\tau$}\mid \mbox{\boldmath $z$}) $;  the matrices
   $\mbox{\boldmath $C_n$}$ and  $\mbox{\boldmath $C_s$}$ are also given, but 
were omitted for brevity.  

One can  use Bayes formula to incorporate the prior knowledge on message and noise and  write the adequate posterior probability:
\begin{eqnarray}
\label{eq:problem}
 P(\mbox{\boldmath $S$},\mbox{\boldmath $\tau$}\mid \mbox{\boldmath $z$})=
\frac{1}{Z} \;\chi\left\{ \mbox{\boldmath $C_s$}\mbox{\boldmath $S$} + 
\mbox{\boldmath $C_n$}\mbox{\boldmath $\tau$}=\mbox{\boldmath $z$} \mbox{ (mod } 2)\right\} P(\mbox{\boldmath $S$}) P(\mbox{\boldmath $\tau$}),
\end{eqnarray}
where the indicator function is  $\chi\left\{A \right\}=1$ if $A$ is true and $0$ otherwise.

 The matrices are chosen at random in such a way that
$\mbox{\boldmath $C_n$}$ is invertible over the field $\{0,1\}$ and  a 
row $m$ in $\mbox{\boldmath $C_s$}$ and 
 $\mbox{\boldmath $C_n$}$ contains  
$K_m$ and $L_m$ non-zero  elements respectively. 
In the same way, each column $j$ of $\mbox{\boldmath $C_s$}$
contains $C_j$ non-zero elements and each column $l$ of 
$\mbox{\boldmath $C_n$}$ contains $D_l$ non-zero elements.

Parity-checks for the signal and noise bits are specified by  the 
matrices $\mbox{\boldmath $C_s$}$ and $\mbox{\boldmath $C_n$}$ respectively. 
The system can be mapped onto a  bipartite graph  represented  by 
$(\mbox{\boldmath $C_s$}\mid\mbox{\boldmath $C_n$})$ (adjacency matrix in 
the graph theory jargon), to say,
 each one of the $M$ rows lists the bit nodes connected to a  check node 
 and each one of the $N+M$ columns lists the checks conveying information 
 about the particular  bit node. Therefore, the sets $\{K_m\}_{m=1}^M$ and
 $\{L_n\}_{n=1}^M$ give the order of check nodes, $\{C_j\}_{j=1}^N$
 and  $\{D_l\}_{l=1}^M$  the order of bit nodes.
 Clearly this sets must obey the relations:  
 \begin{equation}
\sum_{j=1}^{N}C_j=\sum_{m=1}^{M}K_m 
\qquad \sum_{l=1}^{M}D_l=\sum_{m=1}^{M}L_m,
\end{equation}
standing for the number of edges in the signal and noise 
graphs respectively.

The information rate of the code  is given by $R=H_2(p) \;M/N$, where $H_2(p)=-p\mbox{ log}_2(p)-(1-p)\mbox{ log}_2(1-p)$ is the binary entropy of 
the source.

Alternatively one can write $R=H_2(p) \overline{K}/\overline{C}$, where: 
\begin{equation}
\overline{K}=\frac{1}{M}\sum_{m=1}^M K_m \qquad 
\overline{C}=\frac{1}{N}\sum_{j=1}^N C_j 
\end{equation}

To simplify the calculations we change, as in the original work by 
Sourlas \cite{sourlas},
the representation of the variables, replacing the field $\{0,1\}$ 
by $\{\pm 1\}$ and modulo 2 sums by products. 
Moreover, we restrict our analysis to the case of irregular 
bit nodes (sets  $\{C_j\}_{j=1}^N$ and  $\{D_l\}_{l=1}^M$) and regular 
check nodes (fixed $K$ and $L$). The case with regular bit nodes and irregular check nodes is the basis for  high performance codes studied in \cite{IKS}.

\section{Equilibrium theory}

 \label{section3}
To assess the performance of irregular MN codes we compute, using standard 
techniques, the free energy of the system
$f=-\mbox{lim}_{N\rightarrow \infty} \frac {1}{N} \langle \mbox{ln }Z
\rangle$  where $Z$ is the normalisation in (\ref{eq:problem}). The average
$\langle ...\rangle$ is performed over the matrices $\mbox{\boldmath $C_n$}$ and  $\mbox{\boldmath $C_s$}$, the messages $\mbox{\boldmath $\xi$}$ and the noise $\mbox{\boldmath $\zeta$}$ and  will provide information  about the {\it typical} performance of these codes. 

In the ${\pm 1}$ representation,  the syndrome vector  $\mbox{\boldmath $z$}=\mbox{\boldmath $C_n$}\mbox{\boldmath $r$}=\mbox{\boldmath $C_s$}\mbox{\boldmath $\xi$}+\mbox{\boldmath $C_n$}\mbox{\boldmath $\zeta$}\mbox{ (mod 2)} $ 
becomes  ${\cal J}_{\mu\sigma}=
\prod_{j\in\mu} \xi_j \prod_{l\in\sigma} \zeta_l$,
 where $\mu=\langle i_1,\cdots,i_K\rangle$ and 
$\sigma=\langle l_1,\cdots,l_L\rangle$ are sets of indices corresponding to 
the non-zero elements in one of the $M$ rows of $\mbox{\boldmath $C_s$}$ and 
$\mbox{\boldmath $C_n$}$ respectively.
  
The prior distribution over the message bits $S_j\in\{\pm 1\}$  becomes  
$P(S_j)=(1-p)\,\delta(S_j-1)+p\,\delta(S_j+1)$, while 
for the  noise bits  $\tau_l\in\{\pm 1\}$ one has  
$P(\tau_l)=(1-f)\,\delta(\tau_l-1)+ f\,\delta(\tau_l+1)$.

The code construction  is  specified by the tensor 
${\cal A}_{\mu\sigma}\in \{0,1\}$ that determines the
set of indices $\mu\sigma$ which correspond to
non-zero elements in a particular row of the matrix 
$(\mbox{\boldmath $C_s$}\mid\mbox{\boldmath $C_n$})$.
To cope with  non-invertible  $\mbox{\boldmath $C_n$}$ matrices 
one can start by considering an ensemble with uniformly generated
$M\times M$ matrices. The non-invertible instances can then be made 
invertible by eliminating a $\epsilon\sim{\cal O}(1)$ number of rows 
and columns, resulting in an ensemble of $(M-\epsilon)\times(M-\epsilon)$
invertible $\mbox{\boldmath $C_n$}$ matrices and 
$(M-\epsilon)\times(N-\epsilon)$  $\mbox{\boldmath $C_s$}$ matrices.
As we are interested in the thermodynamical limit we can neglect ${\cal O}(1)$
differences  and compute the averages in the original space of 
$M\times M$ matrices. The averages are then performed over an ensemble of 
codes generated as follows: 
\begin{enumerate}
\item sets of numbers $\{C_j\}_{j=1}^{N}$ and $\{D_l\}_{l=1}^{M}$ are sampled 
independently from distributions ${\cal P}_C$ and ${\cal P}_D$ respectively; 
\item  tensors ${\cal A}_{\mu\sigma}$ are generated such that 
 $ \sum_{\mu\sigma} {\cal A}_{\mu\sigma} = M$, 
 $\sum_{\{\mu:j\in\mu\}}{\cal A}_{\mu\sigma}=C_j$ and 
$\sum_{\{\sigma:l\in\sigma\}}{\cal A}_{\mu\sigma}=D_l$, where 
$\{\mu:j\in\mu\}$ stands for all sets of indices that contain $j$. 
\end{enumerate}

The indicator $\chi$ in (\ref{eq:problem}) can  be replaced by  
 a  more tractable function  that is $E(\mbox{\boldmath $S$},\mbox{\boldmath $\tau$};{\cal A})= 1$, if the dynamical variables $\mbox{\boldmath $S$}$ and 
$\mbox{\boldmath $\tau$}$ satisfy  ${\cal J}_{\mu\sigma}=
\prod_{j\in\mu} S_j \prod_{l\in\sigma} \tau_l$  and  $E(\mbox{\boldmath $S$}, \mbox{\boldmath $\tau$}; {\cal A})=0$ otherwise. 
This  function has the form:
 \begin{equation}
 \label{eq:caracteristica}
 \fl E(\mbox{\boldmath $S$}, \mbox{\boldmath $\tau$}; {\cal A}) =\lim_{\beta\rightarrow \infty} \mbox{exp}\left\{-\beta
 \sum_{\mu\sigma} {\cal A_{\mu\sigma}} \;\left[
 {\cal J}_{\mu\sigma}\prod_{j\in\mu} S_j \prod_{l\in\sigma} \tau_l - 1\right]\right\}.
 \end{equation}

The priors over message and noise take the form of external fields in the statistical physics framework and can be written in an exponential form with
the normalisation incorporated in the partition function $Z$:
\begin{equation}
P(\mbox{\boldmath $S$}, \mbox{\boldmath $\tau$})\sim \mbox{exp}
\left(F_s\sum_{j=1}^N S_j\;+\;F_n\sum_{l=1}^M \tau_l\right),
\end{equation}
the fields are then $F_s=\mbox{atanh}(1-2p)$ and $F_n=\mbox{atanh}(1-2f)$.

As in \cite{KMS,MKSV}, the partition function becomes:
\begin{equation}
\label{eq:partition}
\fl Z=\lim_{\beta\rightarrow\infty}  \mbox{Tr}_{\mbox{\boldmath $S$}, \mbox{\boldmath $\tau$}}
\; \mbox{exp}\left[\beta\left(\sum_{\mu\sigma}
{\cal A}_{\mu\sigma}\left({\cal J}_{\mu\sigma}\prod_{j\in\mu}S_j 
\prod_{l\in\sigma}\tau_l - 1\right) +\frac{F_s}{\beta}\sum_{j=1}^N  S_j + \frac{F_{\tau}}{\beta} \sum_{l=1}^M  \tau_l  \right) \right].
\end{equation}

Performing the gauge transformation $S_j\mapsto \xi_j S_j$ 
and $\tau_l\mapsto\zeta_l \tau_l$  one obtains:
\begin{equation}
\label{eq:hamiltonian}
\fl {\cal H}= -\sum_{\mu\sigma} {\cal A}_{\mu\sigma}\left(\prod_{j\in\mu}S_j \prod_{l\in\sigma}\tau_l - 1\right) -\frac{F_s}{\beta}\sum_{j=1}^N \xi_j S_j - 
\frac{F_{\tau}}{\beta} \sum_{l=1}^M \zeta_l \tau_l.
\end{equation}

The resulting Hamiltonian represents a multi-spin ferromagnet in a  
random field, the disorder is transformed as ${\cal J}_{\mu\sigma}
\mapsto 1$ under the gauge transformation, and therefore,
 is trivial and there is no frustration in the system.  The different phases 
that will appear are then due to competition between the local fields 
and ferromagnetic  interactions. Due to the structure of (\ref{eq:caracteristica}) all the  thermodynamics is obtained in the {\it zero temperature}
limit unlike the Sourlas' code case where  optimal decoding must be carried out  at finite temperatures \cite{KS,VSK,sourlas2,rujan,nishimori,iba}.

The free energy $f(p,f,\alpha,{\cal P}_C,{\cal P}_D)=-\mbox{lim}_{N\rightarrow \infty} \frac {1}{N} \langle \mbox{ln }Z
\rangle_{\mbox{$\cal A$},\mbox{\boldmath $\xi$}, \mbox{\boldmath $\zeta$}}$ can be determined using the replica method  along the same lines as reported 
in \cite{KS,VSK,KMS}, but for 
the irregular case it also depends  on the probability distributions 
${\cal P}_C$ and ${\cal P}_D$ used to generate the ensemble of  codes. The auxiliary variables $q_{\alpha_1\cdots\alpha_m}= N^{-1}\sum_j Z_j S_j^{\alpha_1}\cdots S_j^{\alpha_m}$ and $r_{\alpha_1\cdots\alpha_m}= M^{-1}\sum_l Y_l \tau_l^{\alpha_1}\cdots\tau_l^{\alpha_m}$, and their conjugates
$\widehat q_{\alpha_1\cdots\alpha_m}$ and $\widehat r_{\alpha_1\cdots\alpha_m}$
, emerge from the calculation. The  replica symmetry assumption is  enforced by using the 
ans\"atze:

\begin{equation}
q_{\alpha_1 \cdots \alpha_m}=\int dx \; \pi(x) \;x^m  \qquad 
\widehat{q}_{\alpha_1 \cdots \alpha_m}=\int d\widehat{x} 
\; \widehat{\pi}(\widehat{x}) \;\widehat{x}^m
\end{equation}
and
\begin{equation}
r_{\alpha_1 \cdots \alpha_m}=\int dy \rho(y) \;y^m\qquad 
\widehat{r}_
{\alpha_1 \cdots \alpha_m}=\int d\widehat{y} 
\; \widehat{\rho}(\widehat{y}) \;\widehat{y}^m.
\end{equation}

 The expression for the 
free energy  then follows:
\begin{eqnarray}
\label{eq:freeenergy}
\fl f(p,f,\alpha,{\cal P}_C,{\cal P}_D)=\mbox{Extr}_{\{\widehat{\pi},\pi,
\widehat{\rho},\rho\}} \biggl\{ \alpha \;\mbox{ ln }2 
\\
\lo -\alpha \; \int \left[ \prod_{j=1}^K dx_j \pi(x_j)\right] 
                 \left[ \prod_{l=1}^L dy_l \rho(y_l)\right] 
       \mbox{ ln}\left(1+\prod_{j=1}^K x_j \prod_{l=1}^L y_l\right) 
\nonumber\\
\lo +\;\overline{C}\int dx\; \pi(x) \;d\widehat{x}\; \widehat{\pi}(\widehat{x})
  \mbox{ ln}\left(1+x\widehat{x}\right)
+\;\alpha\;\overline{L}\int dy\; \rho(y) \;d\widehat{y}\; \widehat{\rho}
(\widehat{y}) \mbox{ ln}\left(1+y\widehat{y}\right)
\nonumber\\ 
\lo - \sum_C \; {\cal P}_C(C) \; \int \left[ \prod_{j=1}^C d\widehat{x}_j 
\;\widehat{\pi}(\widehat{x}_j)\right] 
 \left \langle \mbox{ ln}\left[e^{\xi F_s} \prod_{j=1}^C (1+\widehat{x}_j)+ 
e^{-\xi F_s} \prod_{j=1}^C (1-\widehat{x}_j)\right] \right \rangle_{\xi}
\nonumber\\
\lo -\;\alpha\; \sum_D \; {\cal P}_D(D) \; \int \left[ 
\prod_{l=1}^D d\widehat{y}_l \; \widehat{\rho}(\widehat{y}_l)\right] 
 \left \langle \mbox{ ln}\left[e^{\tau F_{\tau}} \prod_{l=1}^D
 (1+\widehat{y}_l)+ e^{-\tau F_{\tau}} \prod_{l=1}^D
 (1-\widehat{y}_l)\right] \right \rangle_{\tau}
\biggr\}\nonumber,
\end{eqnarray}
where $\alpha=M/N=\overline{C}/\overline{K}$.

The system's  states  are  obtained by the extremization above, 
resulting in the saddle-point equations :
\begin{eqnarray}
\label{eq:fields}
\fl \widehat{\pi}(\widehat{x})=\int\prod_{j=1}^{K-1}dx_j\;\pi(x_j)
\prod_{l=1}^{L}dy_l\;\rho(y_l) \; \delta\left[\widehat{x} \;-\;
\prod_{j=1}^{K-1}x_j \prod_{l=1}^{L}y_l\right],\\
\fl \widehat{\rho}(\widehat{y})=\int\prod_{j=1}^{K}dx_j\;\pi(x_j)
\prod_{l=1}^{L-1}dy_l\;\rho(y_l) \; \delta\left[\widehat{y} \;-\;
\prod_{j=1}^{K}x_j \prod_{l=1}^{L-1}y_l\right],\nonumber\\
\fl \pi(x)=\sum_C \frac{C}{\overline{C}} {\cal P}_C(C) 
\int\prod_{j=1}^{C-1}d\widehat{x}_j\;\widehat{\pi}(\widehat{x}_j)
\left\langle \delta \left[x \;-\; \mbox{tanh}
\left(  F_s\xi + \sum_{l=1}^{C-1}\mbox{atanh }(\widehat{x}_l) \right)\right]
\right\rangle_\xi,\nonumber\\
\fl \rho(y)=\sum_D \frac{D}{\overline{D}} 
{\cal P}_D(D) \int\prod_{l=1}^{D-1}
d\widehat{y}_l\;\widehat{\rho}(\widehat{y}_l)\left\langle \delta 
\left[y \;-\; \mbox{tanh}\left(  F_{\tau}\zeta + \sum_{l=1}^{D-1}
\mbox{atanh }(\widehat{y}_l) \right)\right] \right\rangle_\zeta.\nonumber
\end{eqnarray}
The exact meaning of the fields $\pi$, $\widehat{\pi}$, $\rho$ and 
$\widehat{\rho}$ were presented in \cite{VSK,KS0} and will be  further 
discussed in the next section.

Due to (\ref{eq:caracteristica}) the estimate 
for the message is $\mbox{\boldmath $\widehat{S}$}=\mbox{sgn}(\langle \mbox{\boldmath $S$}\rangle_{\beta\rightarrow\infty})$, where the 
average is thermal with Hamiltonian (\ref{eq:hamiltonian}) in the zero
temperature limit.
The decoding performance can be measured by
\begin{equation}
\label{eq:mag}
m=\frac{1}{N}\biggl\langle\sum_{i=1}^N\widehat{S}_i  \xi_i \biggr\rangle_{\mbox{\boldmath $\xi$},\mbox{\boldmath $\zeta$},{\cal A}}= \int dh\; \phi(h) \;\mbox{sgn}(h),
\end{equation}
where, as in \cite{MKSV}
\begin{eqnarray}
\label{eq:campo}
\fl \phi(h)=\sum_C  {\cal P}_C(C) \int\prod_{j=1}^{C}d\widehat{x}_j\;\widehat{\pi}(\widehat{x}_j)\left\langle \delta 
\left[h \;-\; \mbox{tanh}\left(  F_s\xi + \sum_{l=1}^{C}\mbox{atanh }(\widehat{x}_l) \right)\right] \right\rangle_\xi.
\end{eqnarray}

Solutions  can be found easily for the case where  $F_s=0$ (unbiased messages) 
and the code constructions are generated by distributions 
$P_D(D)$ and $P_C(C)$ that vanish for $0\leq C,D <2$ (codes with at least two checks per  bit).
For  $K,L>2$  one finds just two types of solutions:  a  ferromagnetic  state with magnetization $m=1$,
\begin{eqnarray}
\label{eq:ferro}
\pi(x)=\delta[x-1] &\qquad \widehat{\pi}(\widehat{x})=\delta[\widehat{x}-1]\\
\rho(x)=\delta[y-1] &\qquad \widehat{\rho}(\widehat{y})
=\delta[\widehat{y}-1],\nonumber
\end{eqnarray}
and a paramagnetic state with $m=0$, 
\begin{eqnarray}
\label{eq:para}
\pi(x)=\delta[x] &\qquad \widehat{\pi}(\widehat{x})=\delta[\widehat{x}]\\
\rho(x)=\left\langle\delta[y-\mbox{tanh}(\zeta F_{\tau})]\right\rangle_{\zeta}
 &\qquad \widehat{\rho}(\widehat{y})=\delta[\widehat{y}].\nonumber
\end{eqnarray}

For other parameter choices, suboptimal ferromagnetic states with $0<m<1$  can also be found by solving the 
saddle-point equations (\ref{eq:fields}) numerically.

The paramagnetic and ferromagnetic free energies can be easily computed 
by inserting (\ref{eq:ferro}) and (\ref{eq:para}) in (\ref{eq:freeenergy})
to give $f_{\mbox{\scriptsize para}}= \alpha\;\mbox{ln }2 -\alpha
\;\mbox{ln ( $2$ cosh }F_{\tau})$ and 
$f_{\mbox{\scriptsize ferro}}=-\;(1-2f)\;F_{\tau}$ respectively. One can
instantly obtain a phase transition occurring at the critical code rate
for the BSC $R_c=1-H_2(f)$, that is valid for every  code 
construction under the restrictions $K,L>2$, $C_j>1$ and $D_l>1$.
This is the same phase transition as the one described in \cite{KMS}.
The critical code rate saturates the channel capacity and
therefore Shannon's coding limit.  

It is important to stress that the coding bound can {\it only} be attained in the case of unbiased messages. For biased messages ($F_s\neq 0$) the paramagnetic state
(\ref{eq:para}) is not a solution for the saddle-point equations (\ref{eq:fields}) and the thermodynamical transition  can only be obtained numerically and 
must be bellow the Shannon's bound as can be shown by a simple upper bound  proposed in \cite{mackay1}.

The upper bound is based on the fact that each bit of the syndrome  
vector  $\mbox{\boldmath $z$}=\mbox{\boldmath $C_n$}\mbox{\boldmath $r$}=\mbox{\boldmath $C_s$}
\mbox{\boldmath $\xi$} + \mbox{\boldmath $C_n$}\mbox{\boldmath $\zeta$}\mbox{ (mod } 2)$ is a sum (or product, depending on the representation adopted) 
 of $K$ message bits with bias  $p$  with $L$ noise bits with flip rate $f$.
The probability of $z_i=+1$ is $p_z^+(K,L)=1/2\;(1+(1-2p)^K(1-2f)^L)$. The maximum information content in the syndrome vector is then $M H_2(p_z^+)$.
For the decoding process one has $M H_2(p_z^+) \ge N H_2(p) + M H_2(f) $, 
resulting in the bound $R\leq  H_2(p_z^+) -  H_2(f) $.  Shannon's  bound
is recovered for  unbiased patterns $p_z^+=1/2$, while for biased patterns
the attainable rates must be bellow Shannon's bound as  $H_2(p_z^+)<1$.

The main question that remains to be addressed is the accessibility of 
the various  states  by a  practical decoding algorithm. In particular, we will focus
on the belief propagation decoding process. In this practical scenario  
the energy landscape may be
dominated by the basin of attraction  of  paramagnetic or suboptimal
ferromagnetic states even when the ferromagnetic state is
the  global minimum, degrading  the practical  performance of the code. 


\section{Statistical physics and belief propagation}

The decoding problem focuses on finding a Bayes-optimal estimate (also known as {\it marginal posterior maximiser}, MPM) 
$\mbox{\boldmath $\widehat {S}$}$ for the original message, given the 
code structure, the syndrome vector  ${\cal J}$ and prior probabilities
 $p$ and $f$. 

The Bayes-optimal estimator is defined as an estimator that minimises the
posterior average of some determined loss function.
Using the overlap  between message and estimate as a loss function, 
the Bayes-optimal 
estimator that emerges is of the form  $\widehat{S}_j=
\mbox{sgn}\langle S_j\rangle_{P(S_j\mid{\cal J})}$ \cite{iba}.
The task of computing this estimator is usually very difficult as no simple form is known for the posterior ${ P}(S_j\mid{\cal J})$ and an exponential
number of operations is required.

\vspace{0.5cm}
\begin{figure}
\hspace*{3cm}
\begin{center}
\epsfig{file=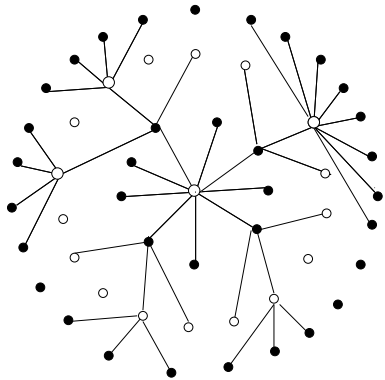,width=50mm}
\vspace{0.2cm}
\caption{Tanner graph representing the neighbourhood of a bit node in an irregular MN code. Black circles represent checks and white circles represent bits.}
\label{figure1}
\end{center}
\end{figure}
 
The problem can be solved in practical time  scales by applying the 
 belief propagation (BP)\cite{pearl} framework. 
In this framework, an approximation
for  the marginal  posterior probabilities $P(S_j\mid{\cal J})$ can be
 computed iteratively in linear time. For that,  a graphical representation (belief network) for 
dependencies between check nodes (or evidence nodes) and signal nodes 
can be constructed. By identifying proper 
substructures in the belief network one can write a closed set of equations 
whose solutions provide the approximation to the posterior probabilities. 
These substructures can be uniquely identified with 
conditional distributions. For LDPCs these probability distributions are:
\begin{eqnarray}
\label{eq:distr}
\fl q^{(S)}_{\mu j} =P
(S_j=S \mid \{ {\cal J}_{\nu\sigma\in {\cal M}_s(j)\setminus \mu} \})
\qquad
\widehat{q}^{(S)}_{\mu j}={P}({\cal J}_{\mu\sigma}\mid S_j=S,\{{\cal J}_{\nu\kappa\neq\mu\sigma}\}) \\
\fl r^{(\tau)}_{\sigma l} ={ P}(\tau_l=\tau\mid\{{\cal J}_{\mu\kappa\in {\cal M}_n(l)\setminus \sigma}\}) 
\qquad
\widehat{r}^{(\tau)}_{\sigma l}={ P}({\cal J}_{\mu\sigma}\mid \tau_l=\tau,\{{\cal J}_{\nu\kappa\neq\mu\sigma}\}),
\end{eqnarray}
where ${\cal M}_s(j)\setminus \mu$ (${\cal M}_n(l)\setminus \sigma$) denote
the set of checks connected to the signal bit $j$ (noise bit $l$) excluding the check  containing the  bits in $\mu$ (noise bits in $\sigma$).
Using Bayes' theorem, the posterior probabilities ${\cal P}(S_j\mid\cal J)$ can then be written in terms of $\widehat{q}^{(S)}_{\mu j}$ and  a priori distributions ${P}_0(S)$ \cite{KS0}.

The Gibbs weight appearing in Equation (\ref{eq:partition}), as observed
 in \cite {sourlas2,KS0}, is proportional to 
${P}({\cal J}\mid \mbox{\boldmath $S$}) {P}_0(\mbox{\boldmath $S$})$
and can be used to write  update formulas for the distributions. 
Introducing $m^s_{\mu j}=q^{(+1)}_{\mu j}-q^{(-1)}_{\mu j}$ and
$m^n_{\nu l}=r^{(+1)}_{\nu l}-r^{(-1)}_{\nu l}$, 
following the steps described in \cite{KS0} one can find the following 
set of equations:
\begin{eqnarray}
\label{eq:belief1}
\fl m^{s}_{\mu l}=\mbox{tanh}\left[ \sum_{\nu\in{\cal M}_s(l)\setminus\mu}
\mbox{atanh}(\widehat{m}^s_{\nu l}) + F_s \right]
\qquad
\widehat{m}^s_{\mu j}={\cal J}_{\mu}\prod_{i\in{\cal L}_s(\mu)\setminus j}m^s_{\mu i} \prod_{l\in{\cal L}_n (\mu)}  m^{n}_{\mu l},
\end{eqnarray}
\begin{eqnarray}
\label{eq:belief2}
\fl m^{n}_{\sigma l}=\mbox{tanh}\left[ \sum_{\nu\in{\cal M}_n(l)\setminus\sigma}\mbox{atanh}(\widehat{m}^n_{\nu l}) + F_n\right]
\qquad
\widehat{m}^n_{\mu j}={\cal J}_{\mu}\prod_{i\in{\cal L}_s(\mu)}m^s_{\mu i} \prod_{l\in{\cal L}_n (\mu)\setminus j}  m^{n}_{\mu l},
\end{eqnarray}
where the set of signal bits (noise bits) in a check $\mu$ ($\sigma$) is 
represented  by ${\cal L}_s(\mu)$ (${\cal L}_n(\mu)$).
The notation ${\cal L}_s(\mu)\setminus l$  indicates all  bits in check $\mu$ excluding  bit $l$, Greek letters run from $1$ to $M$ and Latin letters run from $1$ to $N$.

The  estimate for the message is $\widehat{S}_j=\mbox{sgn}(m^s_j)$, where
$m^s_j$ is computed as:
\begin{equation}
\label{eq:pseudoposterior}
\fl  m^s_{j}=\mbox{tanh}\left[ \sum_{\nu\in{\cal M}_s(j)}
\mbox{atanh}(\widehat{m}^s_{\nu j}) + F_s\right]
\end{equation}

The BP decoding dynamics consists of  updating Equations (\ref{eq:belief1}) and (\ref{eq:belief2}) until a certain halting criteria is reached, and then computing the estimate for the message using equation (\ref{eq:pseudoposterior}). 
The initial conditions are set to reflect the prior 
knowledge about the message  $m^s_{\mu j}(0)=1-2p$ and noise 
$m^n_{\sigma l}(0)=1-2f$.

The BP algorithm is known to provide the {\it exact}  posterior when the 
Tanner graph (see \cite{kschischang}  and references therein)  associated to the system has a tree architecture.
A Tanner graph is a bipartite graph  where checks
are represented by black circles,  bits are represented by white circles and
an edge connects   bits to their related checks.

When very sparse matrices are used, the probability for a loop in the related graph in  a finite number of generations  decays as $\gamma/N$, where $\gamma\sim{\cal O}(1)$ \cite{richardson}. For finite systems one can expect that a limited neighbourhood of node  has a tree structure. When applying the thermodynamical limit $N\rightarrow\infty$, the topology actually converges to a tree and  BP equations become exact. In Figure \ref{figure1} we show a Tanner
graph representing the neighbourhood of a bit node in a large irregular
 MN code.

Equations (\ref{eq:belief1}) and (\ref{eq:belief2}) can also be obtained by looking for extrema of the TAP free-energy \cite{MKSV}:
\begin{eqnarray}
 \fl   f_{\mbox{TAP}}(\mbox{\boldmath $m$}, \mbox{\boldmath $\widehat{m}$})
    &=&\frac{M}{N}\ln 2 +
    \frac{1}{N}\sum_{\mu=1}^M \sum_{i \in {\cal L}_s (\mu)} \ln
    \left(
      1+m_{\mu i}^s \widehat{m}_{\mu i}^s
    \right)
    +\frac{1}{N}\sum_{\mu =1}^M \sum_{j \in {\cal L}_n (\mu)} \ln
    \left(
      1+m_{\mu j}^{n} \widehat{m}_{\mu j}^{n} \right) \nonumber \\
    &\phantom{=}& - \frac{1}{N}\sum_{\mu=1}^M
    \ln \left(
      1+{\cal J}_{\mu}\prod_{i \in {\cal L}_s (\mu)}m_{\mu i}^s
        \prod_{j \in {\cal L}_n (\mu)} m_{\mu j}^{n}\right) \nonumber \\
    &\phantom{=}&- 
    \frac{1}{N}\sum_{i=1}^N \ln
    \left[
      e^{F_s} \prod_{\mu \in {\cal M}_s (i)}
      \left(
        1+\widehat{m}_{\mu i}^s
      \right)
      +e^{-F_s} \prod_{\mu \in {\cal M}_s (i)}
      \left(
        1-\widehat{m}_{\mu i}^s
      \right)
    \right] \nonumber \\
    &\phantom{=}&-
    \frac{1}{N}\sum_{j=1}^M \ln
    \left[
      e^{F_n} \prod_{\mu \in {\cal M}_n (j)}
      \left(
        1+\widehat{m}_{\mu j}^{n}
      \right)
      +e^{-F_n} \prod_{\mu \in {\cal M}_n (j)}
      \left(
        1-\widehat{m}_{\mu j}^{n}
      \right)
    \right] \  .
\label{eq:TAP_freeenergy}
\end{eqnarray}

Observe that the TAP free energy described  above  is not equivalent
to the variational  mean-field free energy introduced in \cite{mackay2,mackay3}.
Here no essential correlations except those related to the presence of loops are disregarded.
 
The meaning of the fields introduced in the previous section 
can be understood by first applying the gauge transformations 
$m^s_{\mu j}\mapsto \xi_j
m^s_{\mu j}$, 
$\widehat{m^s}_{\mu j}\mapsto 
\xi_j\widehat{m^s}_{\mu j}$, 
$m^n_{\sigma l}\mapsto \zeta_l m^n_{\sigma l}$ and
$\widehat{m^n}_{\sigma l}\mapsto\zeta_l \widehat{m^n}_{\sigma l}$ to 
the TAP free energy and  introducing new variables  $x\equiv m^s_{\mu j}$, 
$\widehat{x}\equiv \widehat{m^s}_{\mu j}$,  
$y\equiv m^n_{\sigma l}$ and $\widehat{y}= \widehat{m^n}_{\sigma l}$. 
If $x,\widehat{x},y$ and $\widehat{y}$ are interpreted as random variables
 generated by the probability distributions $\pi,\widehat{\pi},\rho$ and 
$\widehat{\rho}$ respectively, one recovers the replica symmetric free energy (\ref{eq:freeenergy}) (see also \cite{VSK}). 

From the statistical physics point of view, belief propagation is one
of many possible ways to find minima of the TAP free energy, representing
simple iterative fixed point maps. The ferromagnetic state, corresponding to perfect decoding 
is the  global minimum up to Shannon's limit in the case of unbiased
messages (or very close to it in the case of biased messages). However,
this equations are very sensitive to the presence of local minima
in the landscape and the convergence to the global minimum is only expected
if the initial conditions are set up within the basin of attraction of
the ferromagnetic state, which requires prior knowledge about the message 
sent what  is not  the case in practical applications.

In the next sections we will try to address how the free energy 
landscape changes with the parameters.

\vspace{0.8cm}
\begin{center}
\begin{figure}
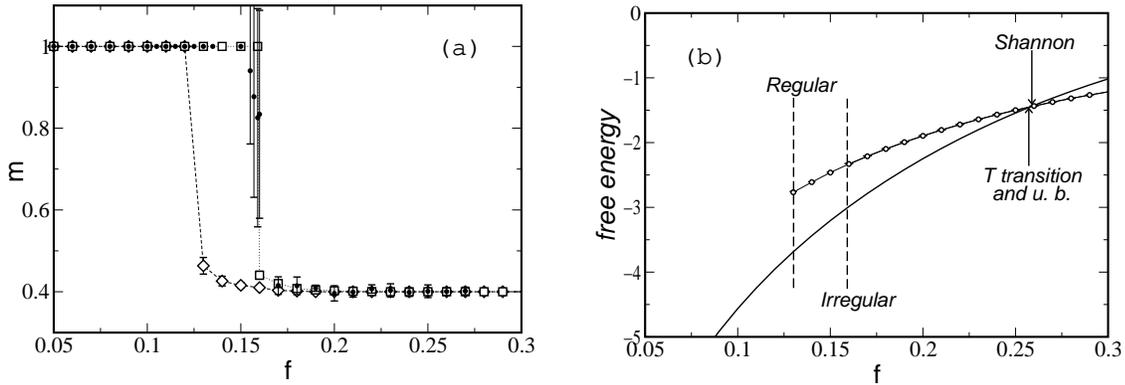

\hspace{0.5cm}

\epsfig{file=mag.eps,angle=-90, width=70mm}\hspace*{0.8cm}\epsfig{file=free.eps,angle=-90,width=70mm}

\caption{(a) Magnetization   as a function of the noise level $f$ for codes 
with $K=L=3$ and $\overline{C}=15$ with message bias $p=0.3$. Analytical RS
solutions for the regular code are denoted as $\Diamond$ and
for the  irregular code; with $C_o=4$ and $C_e=30$ denoted as $\Box$.
 Results are averages over  $10$ runs of the TAP/BP algorithm in an irregular 
code of size $N=6000$ starting from fixed initial conditions (see the text)
; they are plotted as $\bullet$ in the rightmost curve for comparison. TAP/BP results for the regular case agree with the theoretical solutions and have been 
omitted to avoid overloading the figure.
 (b) Free energies for the ferromagnetic state (full line) and for the failure state (line with
$\circ$). The transitions observed in (a) are indicated by the dashed lines.
 Arrows indicate the thermodynamical (T) transition, the upper bound (u.b.)
of Section 3  and Shannon's limit.}
\label{figure3a}
\end{figure} 
\end{center}

\section{Error-correction: regular vs. irregular codes }
Irregularity  improves the practical performance of a MN code. We now illustrate this for the simplest possible irregular constructions with a probability distribution describing connectivities of the signal matrix  $\mbox{\boldmath $C_s$}$ chosen to be:
\begin{equation}
\label{eq:cdistr}
 {\cal P}_C(C) = (1-\theta)\;\delta(C- C_o) \;+\;\theta\;\delta(C-C_e). 
\end{equation} 
The mean connectivity is $\overline{C}=(1-\theta)\;C_o\;+\;\theta\; C_e$ and
$C_o<\overline{C}<C_e$; bits in a group with connectivity $C_o$
will be refered as {\it ordinary} bits and bits in a group with connectivity
$C_e$ as {\it elite} bits. The noise matrix $\mbox{\boldmath $C_n$}$ is chosen to be regular.

To gain some insight on  the  effect of irregularity on solving the 
TAP/BP equations (\ref{eq:belief1}) and (\ref{eq:belief2}) we
 performed several 
runs starting from the fixed initial conditions  $m^s_{\mu j}(0)=1-2p$ and  
$m^n_{\sigma l}(0)=1-2f$  as prescribed in the last section. 
 For comparison  we
also iterated the saddle-point equations (\ref{eq:fields}) obtained in the replica symmetric (RS) theory, setting the initial conditions to be $\pi_0(x)=(1-p)\;\delta(x-m^s_{\mu j}(0))\;+\;p\;\delta(x+m^s_{\mu j}(0))$ and $\rho_0(y)=(1-f)\;\delta(y-m^n_{\sigma l}(0))\;+\;f\;\delta(y+m^n_{\sigma l}(0))$, as suggested from the interpretation of the fields  $\pi(x)$ and $\rho(y)$  in the last section.

In Figure \ref{figure3a} (a) we show a typical curve for the  
magnetization  as a function of the noise level. The RS theory agrees 
very well with TAP/BP decoding results. The addition of irregularity
improves the performance considerably.
In Figure \ref{figure3a} (b) we show the free energies of the two emerging 
states. The free energy for the  ferromagnetic state with magnetization $m=1$ is
 shown as a full line, the  failure state (in Figure  \ref{figure3a}
 (a) with magnetization $m=0.4$) is shown as a line marked with $\circ$. The transitions seen in  Figure \ref{figure3a} (a) are denoted by dashed lines. It is clear
that they are far below the thermodynamical (T) transition, indicating that 
the system becomes trapped in suboptimal states for noise levels $f$
 between the observed transitions and the thermodynamical transition.  The thermodynamical transition coincides with the upper bound (u.b.) in Section 3 and is very close to, but below, Shannon's limit which is shown for comparison. Similar  behaviour has already been observed in regular MN codes with $K=1$ in \cite{MKSV}.   
\vspace{0.5cm}
\begin{figure}
\begin{center}
\epsfig{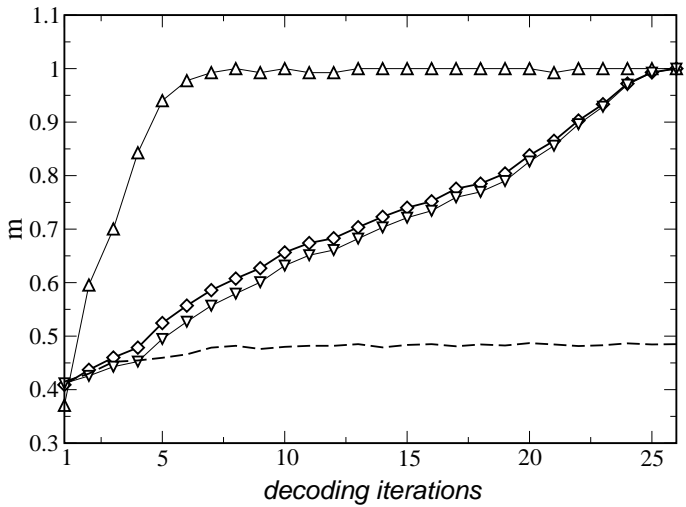}
\end{center}
\vspace*{0.2cm}
\caption{Magnetization monitored during  the TAP/BP decoding  process 
 as a function of the number of  iterations for $N=4000$. Elite nodes 
magnetization is represented by  $\bigtriangleup$. Ordinary nodes 
magnetization is represented by $\bigtriangledown$. The overall magnetization 
is represented by $\Diamond$. The long dashed line shows the dynamics of  the regular code. The constructions employed have  parameters $K=L=3$, $\overline{C}=6$, $C_e=20$ and $C_o=5$. The noise level is $f=0.065$ and the message bias is $p=0.3$.  }
\label{figure3b}
\end{figure}

It is instructive to look how  the magnetization of elite ($m_e$) and
ordinary ($m_o$) nodes evolve throughout  the iterative decoding process. 
In Figure \ref{figure3b} we show this dynamics for a regular and 
an irregular code  at a noise level  where the irregular code converges to
the ferromagnetic state while the regular code fails (long-dashed lines).
One can see that the magnetization of ordinary nodes follow that of the regular code in the first iterations, elite nodes are then corrected quickly achieving high magnetization values. These highly reliable nodes then lead the correction of ordinary nodes (around the fifth iteration), producing successful decoding. From the decoding dynamics point of view  irregular MN codes can be qualitatively regarded  as a mixture of low and highly connected regular codes where elite nodes can tolerate higher noise levels while   ordinary nodes allow for  higher code rates.

\section{The spinodal point} 

In the last section we gained some insight on how irregularity affects the
practical performance of  codes. The dynamical decoding process  shown in 
Figure \ref {figure3b} only provides a qualitative explanation 
and does not seem to allow some simple  analysis.

A possible  alternative is  to relate the observation that the system
gets trapped in suboptimal states (Figure \ref{figure3a}) to global
properties of the free energy. The TAP/BP algorithm can be regarded 
as an iterative solution of fixed point equations for the TAP free 
energy (\ref{eq:TAP_freeenergy}), which  is sensitive to the presence
of local minima in the system. One can expect  convergence to the
global minimum of the free energy  from all initial conditions when there 
is a single  minimum or
when the landscape is dominated by the basin of attraction of this minimum
 when random initial conditions are used.

To analyse this point we rerun the decoding experiments  starting from
initial conditions   $m^s_{\mu j}(0)$ and  $m^n_{\sigma l}(0)$  that 
are random perturbations of the ferromagnetic solution :
\begin{equation}
m^s_{\mu j}(0)= (1-\rho_s)\;\delta(m^s_{\mu j}(0)-\xi_j)\;+\;\rho_s\;\delta(m^s_{\mu j}(0)+\xi_j),
\end{equation}
and 
\begin{equation}
m^n_{\sigma l}(0)= (1-\rho_n)\;\delta(m^n_{\sigma l}(0)-\tau_l)\;+\;\rho_n\;\delta(m^n_{\sigma l}(0)+\tau_l),
\end{equation}
where for convenience we choose $0\leq\rho_s=\rho_n=\rho\leq 0.5$.

We performed TAP/BP decoding several times for different values of $\rho$
and noise level $f$. For $\rho\leq 0.026$ we observed that the system 
converges to the ferromagnetic state for {\em all} constructions, message
biases $p$ and noise levels $f$ examined. It implies that this state is always 
stable. The convergence occurs for any $\rho$ for noise levels below
the transition observed in practice.

\begin{figure}
\hspace*{3cm}
\begin{center}
\epsfig{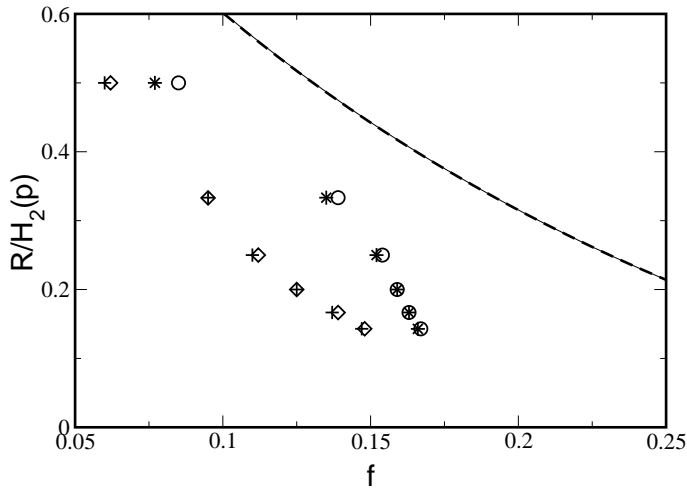}
\caption{Spinodal point noise level $f_s$ for regular and irregular codes.
In both constructions parameters are set as  $K=L=3$. 
Irregular codes with $C_o=4$ and $C_e=30$ are used.
TAP/BP decoding is carried out with $N=5000$ and  a maximum of $500$ 
iterations; they are 
denoted by $+$ (regular) and $\ast$ (irregular). Numerical solutions
for the RS saddle-point equations are denoted by $\Diamond$ (regular) 
and $\mbox{\tiny $\bigcirc$}$ (irregular). Shannon's limit is represented
by a full line and the upper bound in Section 3 is represented by a dashed line. The symbols are chosen to be larger than the actual error bars.}
\label{figureK3L3}
\end{center}
\end{figure}

These observations suggest  that the ferromagnetic basin of
attraction dominates the landscape up to some noise level $f_s$. The fact
that no other solution is ever observed in this region suggests that
$f_s$ is the noise level where suboptimal solutions actually appear, 
namely, it is the noise level that corresponds to 
the {\it spinodal point} of the system. This behaviour have already been observed for regular MN codes with $K=1$ or $K=L=2$ \cite{KMS,MKSV}.

In \cite{KMS,MKSV} we have also shown that MN codes can be divided into three categories with different equilibrium properties: (i)  $K\ge 3$ or $L\ge 3$,
(ii)  $K>1$, $K=L=2$ and (iii)  general $L$, $K=1$. In the next two subsections we will discuss these groups separately.

\subsection { Biased coding: $K\ge 3$ or $L\ge 3$ }

\vspace{0.7cm}
\begin{figure}
\hspace*{3cm}
\epsfig{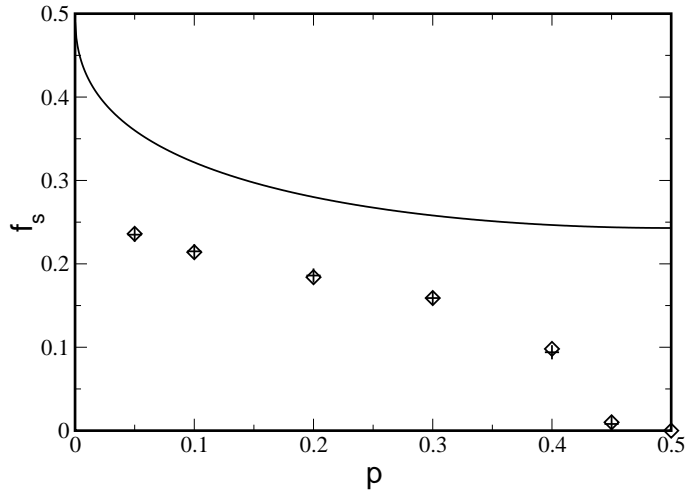}
\caption{Spinodal point $f_s$ for  irregular codes as a function 
of the message bias $p$. The construction is parametrised by 
$K=L=3$, $C_o=4$ and $C_e=30$ with $\overline{C}=15$.
TAP/BP decoding is carried out  with $N=5000$ and a maximum of $500$ 
iterations, and  is represented by $+$, while theoretical RS solutions 
are represented by $\Diamond$. The full  line indicates Shannon's limit.
Symbols are larger than the actual error bars}
\label{figure3}
\end{figure}

To show how irregularity affects codes with this choice of parameters
we choose  $K,L=3$, $C_o=4$, $C_e=30$  and biased messages with  $p=0.3$. 
These choices are arbitrary but can illustrate what happens with the 
practical decoding performance. In Figure \ref{figureK3L3} we show the transition from the  decoding phase to the failure phase as a function of the noise level $f$ for several rates $R$ in  both regular and irregular codes. 
Practical decoding ($\Diamond$ and $\mbox{\tiny $\bigcirc$}$)  results are obtained  for systems of size $N=5000$ with a maximum number of iterations set to $500$. Random initial conditions are chosen and the whole process repeated $20$ times. The practical  transition point is found when the number of failures equals the number of successes.

These experiments were  compared with theoretical values for $f_s$ 
obtained by solving the RS saddle-point equations (\ref{eq:fields})
(represented as $+$ and $\ast$ in Figure \ref{figureK3L3}) and finding the noise level for which a second solution appears. For comparison the coding limit is represented in the same figure  by a full line.

As the constructions used are chosen arbitrarily  one can expect that
these transitions can be further improved, even though the improvement
shown  in Figure \ref{figureK3L3} is already fairly significant.

The analytical solution obtained in Section \ref{section3} for $K\ge 3$ 
or $L\ge3$, $K>1$  and unbiased messages $p=1/2$ implies  that the
system is bistable for  arbitrary code constructions when these parameters
are chosen. The spinodal point noise level is then $f_s=0$ in this case
and cannot be improved by adding irregularity to the construction. Up to
the noise level $f_c$ the ferromagnetic solution is the  global minimum 
of the free energy, and therefore  Shannon's limit is potentially saturated,
however,  the bistability makes these constructions unsuitable for practical 
decoding with a TAP/BP algorithm  when unbiased messages are considered.

The situation improves  when biased messages are used. Fixing the 
matrices   $\mbox{\boldmath $C_n$}$ and  $\mbox{\boldmath $C_s$}$ one can determine how the spinodal point noise level  $f_s$ depends on the bias $p$. In Figure \ref{figure3} we compare simulation results with the theoretical predictions of $f_s$ as a function of $p$. The spinodal point noise level  $f_s$ collapses to zero as $p$ increases towards the unbiased case. It obviously suggests the
use of biased messages for practical use of MN codes with parameters $K\ge 3$ 
or $L\ge3$, $K>1$ under TAP/BP decoding.  

\begin{figure}
\hspace*{3cm}
\epsfig{file=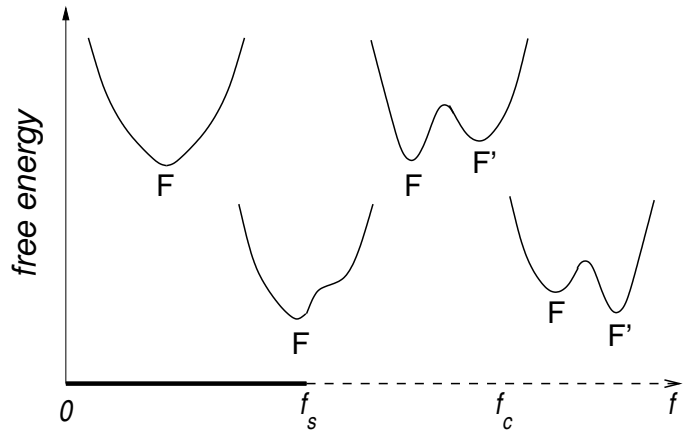,width=90mm}
\caption{Pictorial representation of the free energy landscape as a function
of the noise level $f$. Up to the spinodal point  $f_s$ there is only the
 ferromagnetic state $F$. At $f_s$ another state $F'$ appears dominating the decoding dynamics. The thermodynamical critical noise level $f_c$ indicates the point where the state  $F'$ becomes the global minimum.  }
\label{figure4}
\end{figure}

For biased messages  with $K\ge 3$ or $L\ge3$, $K>1$   the qualitative
picture of the energy landscape differs from the unbiased coding presented 
in \cite{KMS,MKSV}. In Figure \ref{figure4} this landscape  is sketched as a function of the noise level $f$ for
a given bias. Up to the spinodal point $f_s$ the landscape is totally dominated by the ferromagnetic state $F$. At the spinodal point another
 suboptimal state $F'$ emerges, dominating the decoding dynamics. At $f_c$ the suboptimal state $F'$ becomes the global minimum. The bold horizontal line represents
the region where the ferromagnetic solution with $m=1$ dominates the decoding dynamics. In the region represented by the dashed line decoding dynamics  is dominated by 
suboptimal $m<1$ solutions.

\vspace{0.8cm}
\begin{figure}
\hspace*{3cm}
\epsfig{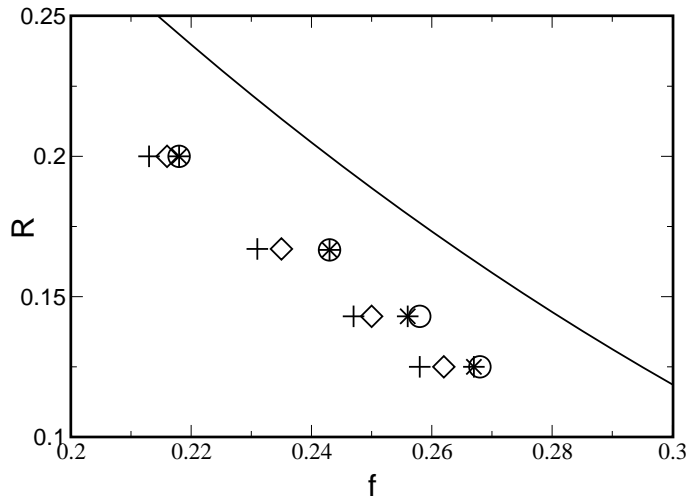}
\vspace{0.3cm}
\caption{Spinodal point noise level $f_s$ for regular and irregular codes.
The constructions are of  $K=1$ and $L=2$, irregular codes are parametrised
 by  $C_o=4$ and $C_e=10$.
TAP/BP decoding is carried out with $N=5000$ and a  maximum of $500$
iterations 
; they are 
denoted by $+$ (regular) and $\ast$ (irregular). Numerical solutions
for RS  equations are denoted by $\Diamond$ (regular)  and 
$\mbox{\tiny $\bigcirc$}$ (irregular). The coding limit is represented by a line. Symbols
are larger than the actual error bars.}
\label{figure5}
\end{figure}

\subsection{Unbiased coding}

For the remaining parameter choices, namely  general $L$, $K=1$ and  $K=L=2$, 
it was shown in \cite{KMS,MKSV} that unbiased coding is generally possible 
yielding close to Shannon's limit performance. The free energy landscape of
 the $K=1$ was shown to behave in a similar way to the one depicted in Figure 
 \ref{figure4} while the landscape  of the case  $K=L=2$ and unbiased messages
  shows a different behaviour where some regions include three stable states  plus their mirror symmetries. 

In the same way as in the $K\ge 3$ case the practical performance is defined
by the spinodal point noise level $f_s$. The addition of irregularity also
changes $f_s$ in these cases.

In the general $L$, $K=1$ family we illustrate the effect of 
irregularity by 
the  choice of $L=2$, $C_o=4$ and $C_e=10$. In Figure \ref{figure5} 
we show the transitions observed by performing $20$ decoding experiments 
with messages of length $N=5000$ and a maximal number of iterations 
set to $500$ ($+$ for regular and $\ast$ for irregular). 
We compare the experimental results with theoretical predictions based 
on the RS saddle-point equations (\ref{eq:fields})
($\Diamond$ for regular and  $\mbox{\tiny $\bigcirc$}$ for irregular).
Shannon's limit is represented by a full line.
The improvement is modest,  what is  expected since regular codes already
present close to optimal  performance. Discrepancies between the 
theoretical and numerical results are due to finite size effects.

\vspace{0.5cm}
\begin{figure}
\hspace*{3cm}
\epsfig{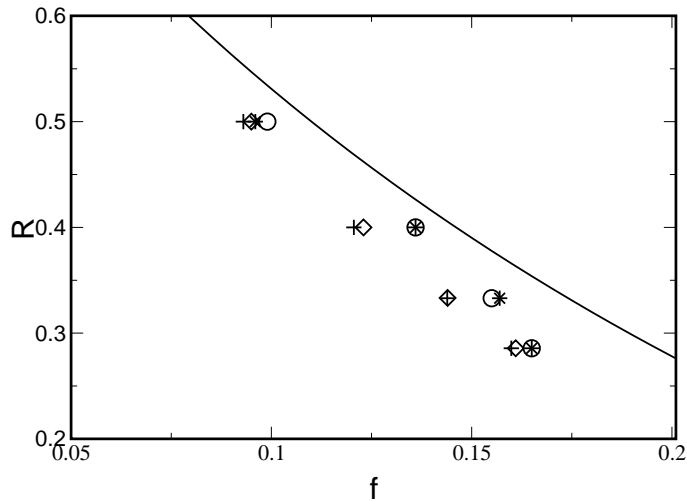}
\vspace{0.2cm}
\caption{Spinodal point noise level values $f_s$ for regular and irregular 
codes.
Constructions are of  $K=2$ and $L=2$, irregular codes are parametrised by
 $C_o=3$ and $C_e=8$.
TAP/BP decoding is carried out with $N=5000$ and a maximum of  $500$ 
iterations; they  are 
denoted by $+$ (regular) and $\ast$ (irregular). Theoretical predictions
 are denoted by $\Diamond$ (regular)  and $\mbox{\tiny $\bigcirc$}$
 (irregular). The coding limit is represented by a line. Symbols are
larger than the actual  error bars.}
\label{figure6}
\end{figure}

We also performed a set of experiments using $K=L=2$ with $C_o=3$ and $C_e=8$,
the same system size $N=5000$ and maximal number of decoding iterations $500$.
The transitions obtained experimentally and  predicted by theory are shown
in Figure \ref{figure6}.

\section{Conclusions} 

We showed that in the thermodynamic limit MN codes are equivalent
to a multi-spin ferromagnet submitted to a random field. 
A replica calculation shows that a phase transition from an {\em errorless} 
(ferromagnetic) phase to a
 {\em failure } (either paramagnetic or suboptimal ferromagnetic) phase  
occurs  as the noise level increases. The phase transition line can be
analytically obtained in the case where constructions with 
$K,L\ge 3$, a minimum of two checks per bit and unbiased messages 
($p=1/2$) are  used. It coincides with Shannon's coding limit and is independent of the code construction.

For other parameter choices the transition only can be obtained numerically 
and coincides with a simple upper bound,  being necessarily below Shannon's limit.

The practical decoding  using belief propagation is shown to attain inferior 
 performance  to Shannon's limit due to the  collapse of the ferromagnetic basin of attraction when new states emerge at the spinodal point noise level $f_s$.
Irregularity increases $f_s$ thus  improving the code's performance. We show
that the maximal noise level corrected by an MN code agrees with the replica theory prediction for the spinodal point noise level $f_s$.

This framework is currently being employed for optimising code constructions
(recently studied in \cite{urbanke}), as well as for finding 
alternatives to the  TAP/BP decoding scheme  and for analysing the 
effect of using inaccurate priors.

\ack
We would like to thank the anonymous referees for their helpful comments.
This work was partially supported by EPSRC grant GR/N00562, a Royal Society travel grant (RV and DS) and by the program ``Research For the Future'' (RFTF)
of the Japanese Society for the Promotion of Science (YK).

\end{document}